\documentclass[letterpaper, 12pt]{article}

\usepackage{setspace}
\usepackage{apacite}

\usepackage{times}

\usepackage{graphicx}
\usepackage{tikz}
\usepackage{amsthm}
\usepackage{amsmath}
\usepackage{natbib}

\usepackage[top=1.2in, bottom=1.2in, left=1.3in, right=1.3in, letterpaper]{geometry}

\newcommand{\X}{\mathbf{X}}
\newcommand{\Score}{\mathbf{S}}
\newcommand{\score}{\mathbf{s}}
\newcommand{\I}{\mathbf{I}}
\newcommand{\V}{\mathbf{V}}
\newcommand{\D}{\mathbf{D}}
\newcommand{\param}{\boldsymbol{\theta}}
\newcommand{\paramhat}{\boldsymbol{\hat\theta}}

\newcommand{\EPC}{\mathrm{EPC}}
\newcommand{\EPCgs}{\EPC_{\mathrm{GS}}}

\begin{document}

\title{The Expected Parameter Change (EPC) for Local Dependence Assessment in Binary Data Latent Class Models}
\author{DL Oberski \and JK Vermunt}
\date{}

\maketitle

\begin{abstract}


Binary data latent class models crucially assume local independence, violations of which can seriously bias the results.
We present two tools for monitoring local dependence in binary data latent class models: the ``Expected Parameter Change'' (EPC) and a generalized EPC, estimating the substantive size and direction of possible local dependencies. 
The asymptotic and finite sample behavior of the measures is studied, and
two applications to the U.S. Census estimation of Hispanic ethnicity and 
medical experts' ratings of x-rays
demonstrate its value in arriving at a model that balances realism and parsimony.

	R code implementing our proposal and including both example datasets is available online as supplementary material. 
	
	\vspace{12pt}
\end{abstract}

\section{Introduction}

The latent class model for binary data is a discrete finite mixture of binomials \citep{agresti_categorical_2002}, and has a wide range of applications in a diverse number of fields. In the social sciences, \citet{hill2001classification} classified patterns of longitudinal change in Swiss voters' support for car pollution abatement policies, while \citet{johnson1990measurement} evaluated the measurement properties of alternative questions to measure ethnicity in the U.S. Census; in machine learning, the model has been used for classifying documents based on word events under the pseudonym ``probabilistic latent semantic analysis'' \citep{hofmann2001unsupervised}; and in education, \citet{dayton1988concomitant} analyzed how elementary school children's ability to correctly answer questions on addition, subtraction, multiplication, and division might indicate mastery of the subject.
In the (bio)medical sciences, latent class analysis (LCA)  for binary data has proved key to describing prevalence and symptomatology of diseases and assessing the accuracy of diagnoses \citep{faraone1994measuring},
and to evaluating the sensitivity, specificity, and predictive validity of diagnostic tests in the absence of a gold standard \citep{walter1988estimation,hui1998evaluation,garrett2002methods}.

The essential assumption in LCA is local independence. One way of viewing the local independence assumption is that it is assumed that, besides the latent class variable, there are no other unobserved variables influencing at least two indicators. For instance, one document ``topic'' may not suffice to explain the number of times pairs of words occur together in it;  addition and subtraction test items may be more strongly associated to one another than to multiplication and division items; and a pair of radiologists might rate x-rays similarly if they trained in the same hospital. For more detailed explanations of how local dependence may arise, we refer to \citet[section 5.2]{biemer2011latent}.
An important distinction to make is then whether this additional unobserved variable is of substantive interest or not \citep{oberski2013local}. For example, an educational researcher may, under certain circumstances, wish to distinguish between ``higher'' and ``lower'' arithmetic skills, whereas 
radiologist judgements' dependence due to having trained in the same hospital is likely to be no more than a nuisance.

Local dependence is a potential problem because it can severely bias LCA outcomes of interest: estimates of classification error, class sizes, and the posterior classification of cases are all affected when local dependence is present \citep{vacek1985effect,torrance1998effects,albert2004cautionary}. 
\citet[p. 610]{hadgu2005evaluation} argued that in the application of LCA to 
 diagnostic tests,  bias  has serious clinical consequences such as an overrated degree of epidemic control, overtreatment, and unrecognized undertreatment. The degree to which such estimates will be biased depends on the size of the local dependence that is being ignored. Large local dependencies  should not be ignored, whereas small local dependencies are not particularly consequential for the outcomes of interest.

Local dependence leads to LCA model misfit, to which the standard reaction is to increase the number of classes \citep[chapter 6]{mclachlan2000finite}. The additional classes then represent (absorb) the dependencies. An alternative method is to model the additional variables explicitly \citep{hagenaars_loglinear_1993,dendukuri2009modeling}. These methods are applicable when the dependence is of substantive interest, but introduce a problem of model interpretability when it is not \citep{yang1997latent,oberski2013local}. 

When local dependence is not of direct interest but represents a nuisance, a more interpretable model may be obtained by modeling the dependence directly. A variety of extended latent class models has been proposed to this end:  additive probability models \citep{harper1972local}; loglinear (logistic) direct effects between indicators \citep{hagenaars1988latent,formann1992linear};  models with   continuous random effects \citep{qu1996random}; and marginal models \citep{bartolucci2006class,reboussin2008locally}. These modeling approaches are in principle attractive when the local dependence is not of substantive interest, but suffer from two problems. First, not all local dependencies may be globally identifiable from the data \citep{jones2010identifiability,stanghellini2013identification}. Second, even when the local dependencies are locally identifiable in some part of the parameter space, the model estimates may become highly unstable; in particular, the substantive latent class variable and the nuisance local dependencies may become difficult to separate. 

Because it is not usually desirable to model all possible local dependencies, the question arises as to \emph{which} pairs of indicators should be modeled as dependent. \citet[800-1]{qu1996random} suggested plotting residual correlations with bootstrapped confidence intervals; \citet[pp. 556-7]{formann2003latent} proposed a hypothesis test on the odds ratio in the residual bivariate crosstables, while \citet[pp. 1063-4]{garrett2000latent} discussed a Bayesian posterior predictive check on the log of that odds ratio; \citet{sepulveda2008biplot} proposed a graphical method based on the log-odds ratios. Finally, \citet[pp. 9--11]{vermunt_latent_2004} proposed using ``bivariate residuals'' (BVR's), the Pearson chi-square in the same crosstables \citep[see also][pp. 72-4]{vermunt_technical_2005}. 
An issue with the BVR, posterior predictive checks, or hypothesis tests is that they focus on the statistical significance of local dependence, while the substantive size of local dependencies is, in our view, the primary motivation for local dependence models: small dependencies are not likely to be relevant for the modeling goals whereas large dependencies should not go undetected. An additional issue with all of these measures is that their development has been \emph{ad hoc}, in the sense that the connection between them and the local dependence model has remained unclear. 

We introduce the ``expected parameter change'' (EPC) measure for detecting local dependencies in latent class models to resolve these problems. The EPC estimates the value that a restricted local dependence parameter would take on if it were freed in an alternative model. 
The EPC therefore directly evaluates the substantive size and direction of possible local dependencies. In addition, the EPC has a clear interpretation in terms of the local dependence model parameterization chosen. For example, when the alternative is the loglinear latent class model with direct effects, the EPC has the interpretation of a conditional log-odds ratio. Other parameterizations are, however, also possible -- including class-specific parameterizations as emphasized by \citet{sepulveda2008biplot}. In this sense the EPC is a generalization of (log) odds ratio residual measures. In addition, a hypothesis test on the EPC is equivalent to a score (``Lagrange multiplier'') test, which in turn was shown to be a generalization of the bivariate Pearson residual (BVR) by \citet{oberski:WP:lca-bvr}.
 The EPC therefore extends these existing local dependence measures while providing a rigorous framework for interpretation in terms of the alternative model. 

We propose two variations of the expected parameter change: the $\EPC_L$ based on the expected information matrix, which is well known in structural equation modeling \citep{saris_detection_1987}, and a novel ``generalized'' $\EPCgs$, based on an information matrix that can be expected to be more robust to model misspecification. 
The $\EPC_L$ is closely related to Rao's classic efficient score test \citep{rao1948large}, while the $\EPCgs$ is related to the generalized score test \citep{white1982maximum,boos1992generalized}.

\vspace{12pt}
The article is organized as follows.
Section \ref{sec:LCM} presents a local dependence latent class model for binary variables. The $\EPC_L$ and $\EPCgs$ for such models are introduced in section \ref{sec:EPC}. Asymptotic and sampling behavior of the $\EPC_L$ and $\EPCgs$ under a range of simulation conditions are then evaluated in section \ref{sec:evaluation}.
In sections \ref{sec:ethnicity} and \ref{sec:dentistry}, two real data applications from the literature, one in the 
social sciences and the other in diagnostic test assessment, demonstrate how these measures can aid in the detection of local dependence and yield different and more easily interpretable results.

\section{Latent Class Model with local dependencies}\label{sec:LCM}

Suppose a sample of size $N$ is obtained on $J$ observed binary variables, aggregated by the $R$ response patterns into $\mathbf{Y}$. 
Let $\mathbf{n}$ be the $R$-vector of observed response pattern counts.
The log-likelihood for 
the latent class model with $T$ classes for the 
unobserved discrete variable $\xi$ can then be formulated \citep{formann1992linear} as the linear-logistic (log linear) model,
\begin{equation}\begin{split}
	\ell(\param) = 	\mathbf{n}' \log \mathrm{Pr}(\mathbf{Y}) 
= \mathbf{n}' \log\left[ \sum^{T}_{t = 1} \mathrm{Pr}(\xi = t) \left( \frac{\exp(\boldsymbol\eta_t)}{\mathbf{1}_R' \exp(\boldsymbol\eta_t)}\right) \right],
\end{split}	\label{eq:loglik}
\end{equation}
where  $\log$ and $\exp$ denote elementwise operations, $\mathrm{Pr}(\xi = t) = \exp(\alpha_t)/\mathbf{1}_T'\exp(\boldsymbol\alpha)$, and 
\begin{equation}
\boldsymbol\eta_t = \X_{(Y)} \boldsymbol{\tau} + 
		\X_{(YY)} \boldsymbol{\psi} + 
		\X_{(Y\xi_t)} \boldsymbol{\lambda}, 
\label{eq:design-matrices}
\end{equation}
where $\X_{(Y)}$, 
 $\X_{(YY)}$ 
and $\X_{(Y\xi_t)}$ 
are design matrices for the observed variables' main effects $\boldsymbol\tau$, bivariate associations $\boldsymbol\psi$, and  associations with the latent class variable $\boldsymbol\lambda$, respectively
\citep{evers1979design}. 
The vector $\boldsymbol\alpha$
contains the logistic main effect parameters for the latent class proportions.
This parameterization of the local dependence latent class model is similar to that adopted
by \citet{hagenaars1988latent} and \citet[section 4.3]{formann1992linear}.

When the observed variables are ``dummy-coded'', the loglinear local dependencies $\boldsymbol{\psi}$ can be interpreted as log-odds ratios between pairs of items within the latent classes. The $\boldsymbol{\psi}$ parameters should in general be interpreted as ``direct effect'' parameters; in effect, they are regression coefficients in the Poisson regression of within-class response pattern counts. Just as in regression analysis, when marginalized over all other response variables, the within-class log-odds dependency between a pair of items may differ from the direct effect parameter $\boldsymbol{\psi}$. Furthermore, even though $\boldsymbol{\psi}$ is equal across classes, the marginal log-odds need not be. The model therefore does allow the marginal log-odds dependence to differ across classes to the extent allowed for by the model. This may be seen as a parsimonious way of modeling dependence. On the other hand, when the goal of the analysis is to interpret the local dependence itself, marginal models such as those discussed by \citet{bartolucci2006class} and \citet{reboussin2008locally} may be more appropriate.

The loglinear parameterization used here is equivalent to a parameterization in terms of probabilities when $\boldsymbol{\psi} = \mathbf{0}$, but has the advantage that probabilities below 0 and above 1 are not possible as sample estimates. For example, consider the $T$-class local independence model for $K$ binary items,
$$
	\mathrm{Pr}(Y_1, \ldots, Y_K) = \sum^T_{t=1}\left[
	\mathrm{Pr}(\xi = t) \prod^{K}_{k=1}
		\mathrm{Pr}(Y_k | \xi = t)
	\right],
$$
in which the joint probability given the latent class $\prod^K_{k=1} \mathrm{Pr}(Y_k | \xi = t)$ may be reparameterized following the usual logistic formulation by taking
$$
\ln \prod_{k=1}^K \mathrm{Pr}(Y_k | \xi = t)	=
\sum_{k=1}^{K} \ln \mathrm{Pr}(Y_k | \xi = t) = 
\sum_{k=1}^{K} \left( \tau_k x_{Y_k}  + \lambda_{kt} x_{Y_k} x_{\xi} \right),
$$
where $x_{Y_k}$ is a design variable that depends on the value of $Y_k$ and $x_{\xi}$ a design variable depending on $\xi$. In dummy coding, for instance, $x_{Y_k}$ will equal 1 if $Y_k$ has the value 1, and 0 otherwise, whereas in effects coding, the corresponding values are +1 and -1. Note that the columns of the design matrices given in Equation \eqref{eq:design-matrices} above are formed by $x_{Y_k}$, $x_{\xi}$, and their products. However, the matrix formulation of the model given above is more flexible as it can also include ``interactions'' (local dependencies) between any number of variables, and can trivially be extended to include covariates predicting class membership. In what follows we  therefore use this convenient formulation of the latent class model.

The $p$-vector of parameters $\param$ can be defined as
$
	\param' = (\boldsymbol{\alpha}', \boldsymbol{\tau}', \boldsymbol{\lambda}', \boldsymbol{\psi}').
$
Thus, the full unconstrained model for binary variables  
has $p = T-1 + J T + {J \choose 2}$ parameters.
Typically, however, not all possible parameters are freed. The standard local independence latent class model, for example, is obtained by setting $\boldsymbol\psi = \mathbf{0}$. More generally, it is also possible to specify parameter restrictions of the  form $\mathbf{a}(\param) = \mathbf{0}$. For the purposes
of this paper, however, we will assume that the restrictions take the form of fixing some or all elements of $\boldsymbol\psi$ to a value (typically zero).

The parameter vector $\param$ can then be partitioned into two parts: a part fixed to a  value and a part corresponding to the  $p$ free parameters of the model. We will denote the fixed  parameter vector by $\param_1$
and the $p$ free parameters by $\param_2$. In the typical latent class
model assuming local independence $\param_1 = \boldsymbol\psi$ and 
$\param_2' = (\boldsymbol{\alpha}', \boldsymbol{\tau}', \boldsymbol{\lambda}')$.

\subsection{Estimation}

The maximum likelihood estimates $\hat\param_2$ under the restricted model can be found by maximizing the likelihood Equation \eqref{eq:loglik} with respect to $\param_2$ while keeping 
$\param_1$ fixed at $\hat\param_1$.   In the local independence model, $\hat\param_1 = \hat{\boldsymbol{\psi}} = \mathbf{0}$. 
Different methods of maximizing Equation \eqref{eq:loglik} have been
suggested in the literature, largely falling into the categories of expectation-maximization on the one
hand \citep{dempster1977maximum} and (quasi-) Newton optimization on the other. Either of these methods or a combination of them may be used to obtain the local maximum when it exists \citep{vermunt2013technical}. 
Since the optimization method used is inconsequential for our following discussion,
we will simply assume that the maximum likelihood 
estimates $\hat\param_2$ can be obtained by one or a combination of these methods. 

\subsection{Identifiability of local dependence parameters}

Local identifiability is a crucial issue for the interpretation of latent class model results and the validity of asymptotic approximations \citep{forcina2008identifiability}. A model is said to be locally identifiable in an open neighborhood when, within this neighborhood, there is one unique set of parameter values that can generate a given likelihood \citep[e.g.][chapter 5]{skrondal2004generalized}.

\citet{goodman1974exploratory} noted that the latent class model will be locally identifiable if the Jacobian, $\Score$, of the expected response patterns with respect to the parameters is of full 
column rank (see also \citealt[Theorem 1]{mchugh1956efficient}; \citealt[Theorem 1]{catchpole1997detecting}; \citealt[p. 1378]{bandeen1997latent}; \citealt{huang2004building}). 
The appendix gives the precise form of this Jacobian for the latent class model in  Equation  \eqref{eq:loglik} with possible local dependencies.
A necessary but not sufficient condition for identifiability is that there should be a nonzero number of degrees of freedom. That is, the number of parameters (columns of $\mathbf{S}$) should not exceed the number of unique patterns (rows of $\mathbf{S}$),  i.e., $p \leq R-1$.

Local identifiability of dependencies is of especial importance for the EPC measures.
The expected parameter change measures to be developed here are  valid only when the single hypothetical local dependency under investigation would be identifiable from the data in the neighborhood of the maximum-likelihood solution.  In general, additional parameters are not necessarily identifiable even when there are positive degrees of freedom. Notwithstanding this general situation, however, for the class-independent local dependencies considered in model \eqref{eq:design-matrices}, it can be proved that identification is possible as long as there are positive degrees of freedom (Theorem 1).
\newtheorem{theorem}{Theorem}
\begin{theorem}\label{theorem}
Consider the model in Equation \eqref{eq:design-matrices} with local dependencies fixed to zero, $\boldsymbol{\psi} = \textbf{0}$. Assume this model is locally identifiable and the number of degrees of freedom is positive, $d > 0$. Then a model including at most $d$ free elements of $\boldsymbol{\psi}$ is locally identifiable.
\end{theorem}
\begin{proof}
	The proof can be found in Appendix A.
\end{proof}
Theorem 1 is useful for the development of the EPC measures, since it greatly simplifies the definition of situations in which EPC measures are appropriate.

\begin{table}[tb]\begin{center}\caption{Number of identifiable local dependence parameters out of
total possible.}\label{tab:identification}
\begin{tabular}{llrrrr}\hline
&	&			\multicolumn{4}{c}{\emph{Number of observed variables ($J$)}}\\\cline{3-6}
&	& $J=3$ & $J=4$ & $J=5$ & $J=6$ \\ \hline
	\multicolumn{2}{l}{\emph{Number of classes ($T$)}} & \\
	&$T=2$ & 0/3 	& 6/6 & 10/10 	& 15/15 \\
	&$T=3$ & -	  	& - &	10/10	& 15/15\\
	&$T=4$ & -		& - &	8/10		& 15/15\\
	&$T=5$ & -		& - &	-		& 15/15\\
	&$T=6$ & -		&  - &	-		& 15/15\\
	\hline
\end{tabular}\end{center}
\end{table}

To demonstrate this result, Table \ref{tab:identification} shows local identifiability of models with an increasing number of classes and variables assessed by the method of \citet[p. 5266]{forcina2008identifiability}. 
Identifiability is examined empirically by randomly sampling a large number of parameter sets and examining the  rank of the expected information matrix for each set. If for each of these random points the information matrix is numerically of full rank, then the model is locally identified with probability close to one \citep{formann1992linear}. 
Shown are the number of local dependencies (elements of $\boldsymbol{\psi}$) that can be identified, where a dash indicates that even the local independence model is not identifiable. Table \ref{tab:identification} shows Theorem \ref{theorem} in action: for instance, since the local independence model with four classes and five response variables is identifiable and has eight degrees of freedom,  exactly eight out of the ten pairwise local dependencies are identifiable.

As shown in model \eqref{eq:design-matrices}, we only consider loglinear local dependencies that are constant over (do not interact with) classes. For loglinear local dependency parameters that may differ over classes, as are considered in graphical models, identifiability conditions are less straightforward. \citet{stanghellini2013identification} provide such conditions for the two-class model, as well as the subspace of parameters in which local identifiability occurs. \citet{jones2010identifiability} investigated identifiability of class-dependent parameters for a particular set of models with covariates. In such cases the derivation of the EPC measures given below will also be applicable, but identification needs to be assessed more carefully.

\section{Expected Parameter Change (EPC)}\label{sec:EPC}

Our approach to monitoring possible local dependencies in latent class analysis sets out from the observation that  local dependencies that have not been parameterized will constitute model misspecifications in the restriction $\boldsymbol{\psi} = \mathbf{0}$. Assuming the local dependencies would be identifiable from the data if parameterized, the expected parameter change ($\EPC$) is an approximately consistent estimate of local dependence misspecifications that can be obtained after fitting the restricted model.
In this section we derive the $\EPC$ and the closely related score test for detecting local dependencies, following the literature on the EPC for structural equation models \citep{saris_detection_1987,sorbom1989model}, and on generalized score tests \citep{boos1992generalized}.
The appendix provides  the first and second derivative 
matrices used in this section.

After estimation of the local independence latent class model, sample estimates $\hat{\boldsymbol{\theta}}$ are obtained that we will assume converge in probability to a population value $\boldsymbol{\theta}^*$ as sample size increases, i.e. $\hat{\boldsymbol{\theta}} \rightarrow \boldsymbol{\theta}^*$. These estimates can be  seen as having been obtained under the model described in the previous section, but with the restriction that all local dependencies are exactly zero, $\boldsymbol{\psi} = \boldsymbol{0}$: this local independence model is  the null model. 

Consider the alternative model that one local bivariate dependence parameter, i.e. some element $\psi$ of the vector $\boldsymbol{\psi}$, is nonzero. Hypothetically this additional parameter $\psi$ could be included in the model as a free parameter, and the model re-estimated. More generally, the part of the parameter vector that is free in both the null and this alternative model is denoted $\boldsymbol{\theta}_2$, while the additional part under consideration such as the local dependence parameter(s) are collected in the vector $\boldsymbol{\theta}_1$. 
The hypothetical parameter estimates that would be obtained under this alternative model are denoted as $\tilde{\boldsymbol{\theta}}$ here. 
Using the standard device of a sequence of local alternatives (e.g.,  \citealt[p. 721]{maydeu2005limited}; \citealt[p. 248]{cameron2005microeconometrics}), the alternative model estimates, as the sample size increases, also converge to the population value $\boldsymbol{\theta}^*$, that is, $\tilde{\boldsymbol{\theta}} \rightarrow \boldsymbol{\theta}^*$. This assumption, common to derivations of the asymptotic distribution of Wald and score statistics, can be informally stated as assuming that the model is ``not too misspecified'', relative to the sample size \citep[e.g.,][p. 6]{kolenikov2010statistics}. Violation of this assumption may occur when the local dependence in question is large and the sample size small, an issue that will be investigated in the simulation study of Section 4.

We now examine the loglikelihood value obtained under the null model, from the point of view of the restriction of interest that the local dependence equals zero. Using a Taylor expansion of the log likelihood at the maximum-likelihood solution,
	\begin{equation}
	\ell \approx \hat{\ell} + \left[\begin{array}{c} 
				\param^*_1 - \hat{\param}_1\\
				\param^*_2 - \hat{\param}_2
	\end{array} \right]'
	\left[\begin{array}{c} 
				\score_1(\param^*)\\ 
				\score_2(\param^*)
	\end{array} \right]
				+ 
		\frac{1}{2}\left[\begin{array}{c} 
				\param^*_1 - \hat{\param}_1\\
				\param^*_2 - \hat{\param}_2
	\end{array} \right]'
	\left[\begin{array}{cc} 
				\I^*_{Y11} & \I^*_{Y21}\\
				\I^*_{Y12} & \I^*_{Y22}
	\end{array} \right]
	\left[\begin{array}{c} 
				\param^*_1 - \hat{\param}_1\\
				\param^*_2 - \hat{\param}_2
	\end{array} \right],
	\label{eq:expansion}
	\end{equation}
where  $\score(\param^*) = \partial \ell / \partial \param$ is the score vector and $\I^*_Y$ is the observed information matrix, both evaluated at the population value $\boldsymbol{\theta}^*$.
As demonstrated in Equation \eqref{eq:expansion}, both are
 partitioned into parameters $\param_2$ included in both the null and alternative models, and parameters $\param_1$ that are being considered as possibly different from their restricted solution,  $\param' = (\param_1', \param_2')$. 

To study what would happen if the restricted parameter vector $\param_1$ were freed, 
we find new estimates by maximizing $\ell$ (Equation \eqref{eq:loglik}), this time with respect to both
$\param_1$ and $\param_2$ \citep[p. 373]{sorbom1989model}. This leads to the equality
\begin{equation}
	\left[\begin{array}{c} 
				\score_1(\param^*)\\ 
				\mathbf{0}
	\end{array} \right] +
	\left[\begin{array}{cc} 
				\I^*_{Y11} & \I^*_{Y21}\\
				\I^*_{Y12} & \I^*_{Y22}
	\end{array} \right]
	\left[\begin{array}{c} 
				\param^*_1 - \hat{\param}_1\\
				\param^*_2 - \hat{\param}_2
	\end{array} \right] = \left[ \begin{array}{c}\mathbf{0}\\\mathbf{0}\end{array} \right].
\end{equation}
Note that $\I^*_{Y}$ cannot be obtained from the maximum likelihood solution as it depends on the unknown value $\param^*$.
However, consistent estimates of the shift in parameter values 
if  $\param_1$ were freed can be obtained from the restricted solution as 
the ``expected parameter change" 
$\EPC =  \hat{\param}_1 - \param_1^* \approx \mathbf{\hat{V}}^{-1} \score_1(\hat{\param})$, 
where $\mathbf{\hat{V}}$ is consistent estimate of $\I^*_{Y}$ evaluated at the 
restricted solution. This implies that $\mathbf{\hat{V}}$ consistently estimates the variance of the score vector $\score_1$, so that a score statistic can be obtained as $T = \score_1(\hat{\param})'\mathbf{\hat{V}}^{-1} \score_1(\hat{\param})$ which is distributed as $\chi^2_{\mathrm{rk}(\Score_1)}$ under the null hypothesis.

Under the null hypothesis $\boldsymbol{\psi} = \mathbf{0}$, the information matrix $\I^*_{Y}$ is consistently estimated by the expected information matrix evaluated at the restricted solution $\hat\I_L$, so that  \citep{rao1948large} 
\begin{equation}\begin{split}
\EPC_L 	&= \hat\V_L^{-1} \score_1(\hat\param) = \hat\I_{L}^{-1}\, \score_1(\hat\param)\\
&=  (\hat{\I}_{L11} - \hat{\I}_{L12}\hat{\I}_{L22}^{-1}\hat{\I}_{L21})^{-1} \,  \score_1(\hat{\param}),
\end{split}\label{eq:epc_l}\end{equation}
where the last step, following from the rules for inverting a partitioned non-singular matrix, is computationally convenient since the partition $\hat{\I}_{L22}$ of the information matrix corresponding to the free parameters will usually already be at hand in latent class modeling software.
The $\EPC_L$ defined above is popular in the field of structural equation modeling  \citep{saris_detection_1987}. \citet{rao1948large}'s efficient score statistic can be obtained as $T_{L} = \score_1(\hat\param) ' \hat\I_{L}^{-1} \score_1(\hat\param)$, which under the null hypothesis has a chi square distribution with $\mathrm{rank}(\Score_1)$ degrees of freedom. The efficient score statistic is known in the structural equation modeling literature
as the ``modification index" (MI) \citep{sorbom1989model}, and in the econometrics literature as the Lagrange multiplier test \citep{aitchison1958maximum,breusch1980lagrange}.
By the same argument of consistency under the null hypothesis, the expected information matrix $\hat{\I}_L$ can be replaced by the observed information  evaluated at the restricted solution, $\hat{\I}_Y$ 
\citep[see][]{glas1999modification,linden2010statistical}. 

The derivation of $\hat{\V}$ under the null hypothesis suggests that when $\boldsymbol{\psi} \neq \mathbf{0}$, the $\EPC_L$ is asymptotically biased. 
Under misspecified local independence, a ``generalized'', i.e. robust to misspecification, consistent estimate $\hat{\mathbf{V}}_{\mathrm{GS}}$ of $\V$ can be used
\citep{white1982maximum}.
As shown by \citet[p. 329]{boos1992generalized},
\begin{equation}
\begin{split}
\hat\V_{\mathrm{GS}}	
		&= (\mathbf{1}, - \hat\I_{Y12} \hat\I_{Y22}^{-1}) \hat\D (\mathbf{1}, - \hat\I_{Y12} \hat\I_{Y22}^{-1})'
		\\
		&=  \hat\D_{11} - \hat\I_{Y12} \hat\I_{Y22}^{-1} \hat\D_{12}' - 
		\hat\D_{12} \hat\I_{Y22}^{-1} \hat\I_{Y12}' + 
		\hat\I_{Y12} \hat\I_{Y22}^{-1} \hat\D_{22} \hat\I_{Y22}^{-1} \hat\I_{Y12}',
\end{split}
\end{equation}
where $\D$ is the outer product matrix of first derivatives of the log-likelihood (see appendix) and $\hat\I_Y$ and $\hat\D$ denote  quantities evaluated at the sample 
estimates $\paramhat$ under the restricted model. 
A ``generalized expected parameter change'' $\EPCgs$ is obtained as 
$\EPCgs = \hat\V_{\mathrm{GS}}^{-1} \score_1(\hat\param)$;
the well known generalized score test \citep{white1982maximum} is $T_{\mathrm{GS}} = \score_1(\hat\param)' \hat{\V}_{\mathrm{GS}}^{-1} \score_1(\hat\param) $. 

\subsection{Form of the $\EPC_L$ for a local dependence parameter}

So far we have given the expected parameter change statistics in generality. That is, the equations given above may, in fact, be applied to any restricted parameter, not only to local dependencies. To gain more insight into the use of these statistics for the detection of local dependence, we give here the form of the EPC for detecting local dependencies between two variables $Y_j$ and $Y_{j'}$ in the parameterization of Equation \eqref{eq:design-matrices}.

As noted above, the $\EPC_L$ is defined as minus the derivative of the log-likelihood with respect to the local dependence parameter, premultiplied by the inverse information matrix under the alternative, both evaluated at the restricted solution,
\begin{equation}
	\EPC_L = \left( \I_L  |_{\param=\hat{\param}} \right) 
	\left( \frac{\partial \ell}{\partial \psi_{jj'}} |_{\param=\hat{\param} } \right),
\end{equation}
where $\param = \left(\boldsymbol{\alpha}', \boldsymbol{\tau}', \boldsymbol{\lambda}', \psi_{jj'} \right)'$.
In the derivatives involved in the information matrix $\I_L$ and $\partial \ell /\partial \psi_{jj'}$, a key role is played by the design matrices $\X_{(Y)}$,  $\X_{(Y\xi_t)}$, and  $\X_{(YY)}$. For example, the derivative $\partial \ell /\partial \psi_{jj'}$ is determined by the corresponding column of the design matrix $\X_{(YY)}$: with dummy coding, this column will be a vector that equals 1 for all patterns in which both $Y_j$ and $Y_{j'}$ equal 1, and 0 otherwise; using effects coding, the corresponding values will be 1 and -1. 

The derivative with respect to a hypothetical local dependence parameter between two variables is closely related to the raw residual in the bivariate crosstable between these two variables. Since in the current model all residuals in the bivariate crosstable will be equal to each other in absolute value, we shall consider $r_{11}$, that is, the observed minus the expected number of observations in which the two variables $Y_j$ and $Y_{j'}$ are both equal to 1, i.e., $r_{11} = n_{11} - \hat{\mu}_{11}$, where $n_{kl}$ and $\hat{\mu}_{11}$ respectively denote the observed and expected number of observations in cell $(k,l)$ of the crosstable. 

From the general form of the derivatives, given in the Appendix, it can then be shown that using effects coding,
\begin{equation}
\begin{split}
 \frac{\partial \ell}{\partial \psi_{jj'}}
  &= \mathbf{n}' \sum_{t \in 1..T} \mathrm{Pr}(\xi = t | \mathbf{Y}) [\mathbf{x}_{(y_j y_{j'})} - \mathbf{x}_{(y_j y_{j'})}'\mathrm{Pr}(\mathbf{Y}  | \xi) ] \\
&= \mathbf{n}' [\mathbf{x}_{(y_j y_{j'})} - \mathbf{x}_{(y_j y_{j'})}'\mathrm{Pr}(\mathbf{Y}) ] 
\\
&=\sum_{k=l} (n_{kl} - \hat{\mu}_{kl}) - \sum_{k \neq l} (n_{kl} - \hat{\mu}_{kl})
= 4 r_{11}
\end{split} \end{equation}
where  the second step is due to the fact that $\psi_{jj'}$ is class-independent, and the last step follows because the off-diagonal residuals have a sign opposite to the diagonal residuals. 
If dummy coding is chosen instead of effect coding, $\partial \ell /\partial \psi_{jj'} = r_{11}$. The precise form of the information matrix $\mathbf{I}_L$ is given in the Appendix. 

There is, therefore, a  close relationship between the residual in the bivariate crosstable and the expected parameter change. The EPC equals the raw residual ``divided by'' its variance under the alternative model.  This finding is in close correspondence with the finding of \citet{oberski:WP:lca-bvr} that the score test is closely related to the bivariate residuals.

\section{Asymptotic and finite sample evaluation of expected parameter change}\label{sec:evaluation}

In this section we evaluate both the asymptotic and sampling performance of the suggested $\EPC_L$ and $\EPCgs$ statistics for detecting relevant local dependencies.
The goal of this investigation is to evaluate the feasibility of the EPC as a measure of the size of possible local dependencies. 

EPC measures when applied to population data should approximate the true size of the local dependence. EPC measures calculated on samples should be sample estimates of the population EPC. 
Under different conditions, we examine:
\begin{itemize}
  \item To what extent the \emph{population} EPC corresponds to the true local dependence;
  \item To what extent the average \emph{sample} EPC corresponds to the population EPC.
\end{itemize}
By performing both a population and a finite sample analysis, we can separate 
errors due to the approximation inherent in the EPC on the one hand from  errors due to sampling fluctuations on the other.

\subsection{Setup}	

The population model is specified as a two-class model with five binary indicators
and one local dependence between a pair of indicators.
In our setup, all design matrices in equation \eqref{eq:design-matrices} are chosen 
such that the columns sum to zero (``effect coding''). The intercepts $\boldsymbol{\tau} = \mathbf{0}$, 
the latent class intercept $\alpha = 0.20$, and the ``loadings'' and bivariate local 
dependence are varied across conditions:
	\begin{enumerate}
		\item Local dependence size ($\psi$):
			-0.50 (high-negative), -0.20 (middle-negative) , -0.05 (low-negative), 0 (none), +0.05 (low-positive), +0.20 (middle-positive), 0.50 (high-positive);
		\item Effect of latent variable on indicators ($\lambda$):
			0.5 (medium-low), 0.8 (high).
	\end{enumerate}
A subsequent Monte Carlo simulation crosses these 14 conditions
with sample size,
	\begin{enumerate}
			\item[3] Sample size (nobs):
			128, 256, 512, 1024, 2048.
	\end{enumerate}
We therefore examine the sampling performance of the two statistics for 70
conditions in total.

\begin{figure}\caption{Effect of local dependence. 
	Shown is the conditional probability that an observed variable $Y_j = 1$ 
	given the latent class variable $\xi$ and a different observed variable
	$Y_{j'}$ ($j \neq j'$), for six conditions. (Only conditions 
	with positive slopes are shown here.)}\label{fig:conditions}
	\input{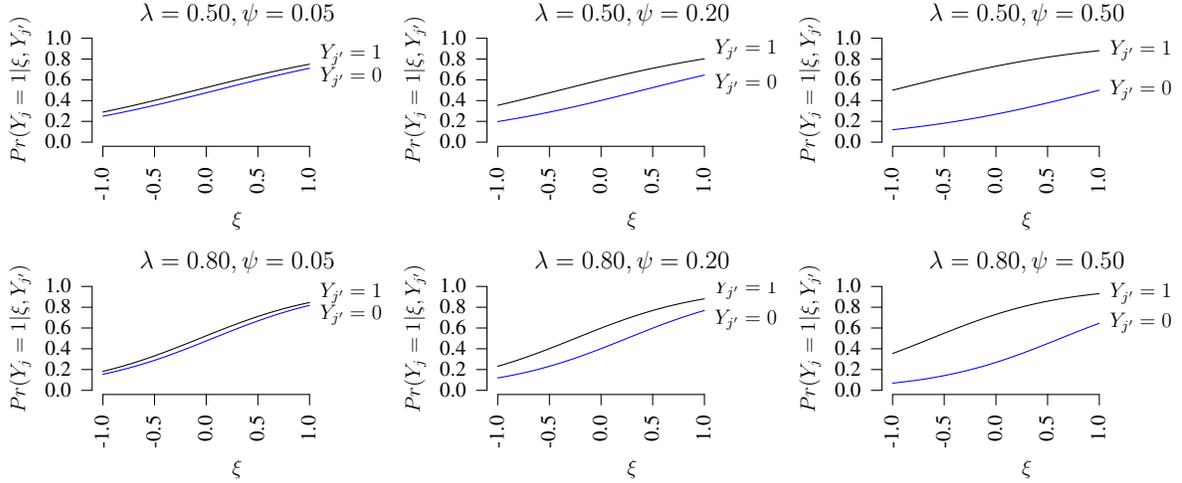}
\end{figure}
	To give the reader an idea of the implications of these conditions, 
	Figure \ref{fig:conditions} shows the
	effect of choosing different combinations
	of the slope parameter $\lambda$ and the local dependence parameter $\psi$ on the 	conditional probability for one observed variable.
	For illustrative purposes, many different values for the latent class 
	variable $\xi$ are plotted; in practice there will be only $T$ points along the 	horizontal axis. The Figure shows that $\psi=0.05$ constitutes
	a rather small local dependency, while choosing $\psi=0.5$ has a very
	large effect on the implied conditional probability. This effect is  
	larger in absolute terms when the slope parameter $\lambda$ is small. 

To illustrate the implications of these conditions, Figure \ref{fig:conditions} depicts how the class-specific response probability for variable $Y_j$ is affected by the value of a different variable $Y_{j'}$ for particular values of $\lambda$ and the local dependence, $\psi$, between the two items. For illustrative purposes, the latent variable $\xi$ is treated as continuous, but in fact it takes on only the values 0 and 1. It can be seen that $\psi=0.05$ constitutes a rather small local dependency (lines are close to one another), while choosing $\psi=0.5$ has a very large effect on the implied class-specific response probabilities. This effect is larger in absolute terms when the slope parameter $\lambda$ is small.

\subsection{Asymptotic performance}

We will first evaluate the asymptotic performance of the $\EPC_L$ and $\EPCgs$ obtained from the $H_0$ model which omits the local dependence. For this purpose, we compute maximum likelihood estimates under the $H_0$ model using the population proportions under the $H_1$ model as data. Since this amounts to minimizing the 
Kullback-Leibler distance, we refer to this model as the ``KL-model''. The KL-model provides the asymptotic value (as the sample size approaches infinity) of the $\EPC$ and score statistic given $H_1$.

The top parts of Tables \ref{tab:epc-L-population} and \ref{tab:epc-GS-population} 
show the obtained $\EPC_L$ and $\EPCgs$ values under the different conditions. 
It can be seen that when there is no misspecification, i.e. when 
the true local dependence parameter is zero, both $\EPC$'s will also estimate zero. 
When there is a small misspecification of -0.05 or +0.05, both EPC's have population 
values that are very close to the true local dependence. 
The top part of Table \ref{tab:epc-L-population} shows that with larger local 
dependencies in absolute value, the population
$\EPC_L$ is a biased estimate of the true local dependence parameter. The 
percentage relative bias in the $\EPC_L$ 
is shown in the bottom part of Table \ref{tab:epc-L-population}.
Local dependencies of +0.2 and +0.5 cause larger asymptotic biases than their
negative counterparts. Under the condition with lower slopes and the
largest positive misspecification, the $\EPC_L$ is an 338\% overestimate
of the true local dependence. In contrast, with negative local dependencies,
the $\EPC_L$ is underestimated in absolute terms. 

\begin{table}[tb]\begin{center}
\caption{Population $\EPC_L$ statistics under the 14 simulation conditions.
}\label{tab:epc-L-population}

\begin{tabular}{rrrrrrrr}
  \hline
	& \multicolumn{7}{l}{Local dependence ($\psi$)}\\\cline{2-8}
$\lambda$	 & -0.5 & -0.2 & -0.05 & 0 & 0.05 & 0.2 & 0.5 \\ 
  \hline
  0.5 & -0.374 & -0.165 & -0.047 & -0.000 & 0.054 & 0.313 & 2.190 \\ 
  0.8 & -0.329 & -0.159 & -0.047 & -0.000 & 0.054 & 0.277 & 1.425 \\ 
  \\
	& \multicolumn{7}{l}{Percent relative bias in $\EPC_L$}\\\cline{2-8}
  0.5 & -25 & -17 & -6 & - & 9 & 56 & 338 \\ 
  0.8 & -34 & -20 & -6 & - & 7 & 39 & 185 \\ 
   \hline
\end{tabular}\end{center}
\end{table}

\begin{table}[tb]\begin{center}
\caption{Population $\EPCgs$ statistics under the 14 simulation conditions.}\label{tab:epc-GS-population}

\begin{tabular}{rrrrrrrr}
  \hline
	& \multicolumn{7}{l}{Local dependence ($\psi$)}\\\cline{2-8}
$\lambda$	 & -0.5 & -0.2 & -0.05 & 0 & 0.05 & 0.2 & 0.5 \\ 
  \hline
  0.5 & -0.403 & -0.186 & -0.050 & -0.000 & 0.050 & 0.181 & 0.694 \\ 
  0.8 & -0.439 & -0.209 & -0.051 & -0.000 & 0.048 & 0.167 & 0.344 \\ 
  \\
	& \multicolumn{7}{l}{Percent relative bias in $\EPCgs$}\\\cline{2-8}
  0.5 & -19 & -7 & -1 & - & -1 & -10 & 39 \\ 
  0.8 & -12 & 5 & 3 & - & -4 & -17 & -31 \\ 
   \hline
\end{tabular}
\end{center}
\end{table}
Table \ref{tab:epc-GS-population} shows the population $\EPCgs$ estimates
(top part) as well as the percentage bias relative to the true population
local dependence (bottom part). The table shows that the relative asymptotic bias
in the $\EPCgs$ is uniformly much lower than that in the $\EPC_L$: on average 
it is 60\% lower. 
Overall the $\EPCgs$ has much better asymptotic performance.

\subsection{Finite sample performance}

In finite
samples, sampling fluctuations in the score and the $\V$ matrix will influence 
the $\EPC$'s as well. We therefore performed a Monte Carlo simulation to evaluate
the sampling behavior of these statistics. 
From each of the 70 populations, a sample
of $N$ observations was drawn and the $\EPC_L$ and $\EPCgs$ were calculated. This
process was replicated 400 times to yield a sampling distribution for 
$\EPC_L$ and $\EPCgs$.

\begin{table}[tb]\begin{center}\caption{Monte Carlo simulation results: median $\EPC_L$ statistics over 400 replications for each condition.
For comparison, the bottom rows provide population values obtained from the KL-model.
}\label{tab:sim-EPC-L}
\begin{tabular}{rrrrrrrrrrr}
  \hline
    	& \multicolumn{8}{l}{Local dependence ($\psi$)}\\\cline{2-9}
  & \multicolumn{2}{c}{$\psi = -0.05$}&& \multicolumn{2}{c}{$\psi = -0.20$}&& \multicolumn{2}{c}{$\psi = -0.50$}\\
    \cline{2-3}  \cline{5-6}  \cline{8-9}
  	& \multicolumn{8}{l}{Loading ($\lambda$)}\\\cline{2-9}
No. obs.  &  0.5 & 0.8 &  &  0.5 & 0.8 &  &  0.5 & 0.8  \\
  \hline
  128  &   -0.053  &  -0.054  &  &   -0.164  &  -0.163  &  &  -0.377  &  -0.330\\
  256  &   -0.040  &  -0.046  &  &   -0.157  &  -0.159  &  &  -0.374  &  -0.333\\
  512  &   -0.055  &  -0.045  &  &   -0.166  &  -0.158  &  &  -0.380  &  -0.332\\
  1024  &   -0.041  &  -0.047  &  &   -0.164  &  -0.160  &  &  -0.378  &  -0.328\\
  2048  &   -0.045  &  -0.051  &  &   -0.163  &  -0.162  &  &  -0.376  &  -0.330\\
  \\
Population  		& -0.047 & -0.047&& -0.165 & -0.159 & & -0.374 & -0.329\\
  \hline
  \\
  & \multicolumn{2}{c}{$\psi = +0.05$}&& \multicolumn{2}{c}{$\psi = +0.20$}&& \multicolumn{2}{c}{$\psi = +0.50$}\\
  \cline{2-3}  \cline{5-6}  \cline{8-9}
No. obs.  &  0.5 & 0.8 &  &  0.5 & 0.8 &  &  0.5 & 0.8  \\
      \hline
  128  &  0.032  &  0.030  &  &  0.212  &  0.235  &  &  1.235  &  0.993\\
  256  &  0.053  &  0.045  &  &  0.278  &  0.282  &  &  1.695  &  1.199\\
  512  &  0.052  &  0.049  &  &  0.282  &  0.292  &  &  1.962  &  1.330\\
  1024  &  0.049  &  0.051  &  &  0.294  &  0.271  &  &  2.079  &  1.358\\
  2048  &  0.055  &  0.057  &  &  0.302  &  0.276  &  &  2.110  &  1.351\\
  \\
Population   & 	0.054 & 0.054 && 0.313 & 0.277 && 2.190& 1.425\\
  \hline
\end{tabular}\end{center}
\end{table}
Table \ref{tab:sim-EPC-L} shows the median $\EPC_L$ estimates over each of the 400 
samples in each of the conditions. For the $\EPC_L$ to be unbiased with
respect to the true local dependence, these values should correspond to the 
size of the $\psi$ local dependence parameter shown in the table headers. 
Considering the population bias reproduced in the rows marked ``population'', 
we would not expect unbiasedness with respect to 
$\psi$ in general. Except in conditions with sample sizes 128 and 256, the median sample estimates in Table \ref{tab:sim-EPC-L} are close to the population values.

With a small sample size of $N=128$, the sample estimates of the $\EPC_L$ are
biased with respect to the population values. Paradoxically, the small-sample 
estimates can be closer to the true misspecification
than the population values are (see for example  the 
conditions with $\psi = +0.20$ and $\psi = +0.50$). As expected, increasing
the sample size brings the median $\EPC_L$ closer to the population value.
It is clear that the conditions with the larger slopes perform much better
than those with lower slopes, both in the population and in finite samples. With 
five indicators and true slopes equal to 0.8, the $\EPC_L$ provides reasonable
estimates in all conditions. Whether this condition is satisfied
cannot be verified from a given restricted sample solution, since the restriction itself
may bias the loading estimates. 

\begin{table}[tb]\begin{center}
\caption{Monte Carlo simulation results: median $\EPCgs$ statistics over
400 replications for each condition. For comparison, the bottom rows provide population values obtained from the KL-model.}\label{tab:sim-EPC-GS}
\begin{tabular}{rrrrrrrrrrr}
  \hline
    	& \multicolumn{8}{l}{Local dependence ($\psi$)}\\\cline{2-9}
  & \multicolumn{2}{c}{$\psi = -0.05$}&& \multicolumn{2}{c}{$\psi = -0.20$}&& \multicolumn{2}{c}{$\psi = -0.50$}\\
    \cline{2-3}  \cline{5-6}  \cline{8-9}
  	& \multicolumn{8}{l}{Loading ($\lambda$)}\\\cline{2-9}
No. obs.  &  0.5 & 0.8 &  &  0.5 & 0.8 &  &  0.5 & 0.8  \\
  \hline
  128  &  -0.052 & -0.053 & & -0.155  & -0.196 & &-0.330 & -0.424 \\
  256  &  -0.040 & -0.049 & & -0.168  & -0.207 & &-0.350 & -0.432 \\
  512  &  -0.057 & -0.051 & &  -0.185  & -0.202 &&-0.375 & -0.439 \\
  1024 &  -0.044 & -0.052 & & -0.186  & -0.208 & & -0.388 & -0.438\\
  2048 &  -0.048 & -0.055 & &  -0.183 & -0.211  && -0.396 & -0.440 \\
  \\
  Population & -0.050 & -0.051 & & -0.186 & -0.209 && -0.403 & -0.439\\
  \hline
  \\
  & \multicolumn{2}{c}{$\psi = +0.05$}&& \multicolumn{2}{c}{$\psi = +0.20$}&& \multicolumn{2}{c}{$\psi = +0.50$}\\
  \cline{2-3}  \cline{5-6}  \cline{8-9}
No. obs. &  0.5 & 0.8 &  &  0.5 & 0.8 &  &  0.5 & 0.8  \\
      \hline
  128  & 0.027 & 0.027 & &  0.102 & 0.136 &	 & 0.305 & 0.214   \\
  256  & 0.047 & 0.042 & &  0.130 & 0.158  &	 & 0.468 & 0.263  \\
  512  & 0.046 & 0.045 & &  0.152  & 0.164 &	 & 0.605 & 0.298 \\
  1024 & 0.044 & 0.046 & &  0.165  & 0.162 &	 & 0.619 & 0.321 \\
  2048 & 0.049 & 0.050 & &  0.171 & 0.166  &	 & 0.670 & 0.326 \\
  \\
  Population & 0.050 & 0.048 && 0.181 & 0.167&& 0.694 & 0.344\\
   \hline
\end{tabular}\end{center}
\end{table}

Table \ref{tab:sim-EPC-GS} shows the Monte Carlo simulation results for 
$\EPCgs$. Even for small sample sizes, the median $\EPCgs$ over simulated samples
is close to the population $\EPCgs$. The sample $\EPCgs$ estimates are close to the 
true local dependence parameters. 
The $\EPCgs$ clearly performs better than 
the $\EPC_L$ both in the population and in finite samples overall. An exception occurs in the case of $\psi = 0.20$ with small samples\footnote{We thank an anonymous reviewer for pointing out this finding.}: here the average $\EPC_L$ values are closer to 0.20 than are the average $\EPCgs$ values, which are downwards biased. This occurs due to the interplay of an upwards bias in the asymptotic value of $\EPC_L$ combined with a downwards small-sample bias. These approximate bias canceling effects only appear to occur in a few circumstances and in practice one may not want to rely on their occurrence, however. 
Overall the bias in the $\EPCgs$ can be viewed as acceptable for the purpose
of detecting the substantive size of local dependencies.

The $\EPCgs$ appears preferable to $\EPC_L$. However, if in practice the observed information matrix is close to singular, $\EPCgs$ will not be computable so that $\EPC_L$ may be an alternative in those cases. On the other hand, when the model is very large, the expected information matrix, which involves all possible response patterns, may require a prohibitively large amount of computer resources; in such cases $\EPC_L$ may not be computable.

\section{Application 1: Measurement of Hispanic ethnicity in the U.S. Census}\label{sec:ethnicity}

\citet{johnson1990measurement} performed a latent class analysis of four indicators of Hispanic ethnicity in the U.S. Census. For 9701 respondents to the 1986 National Content Test,  two  indicators were obtained during an initial interview (at time point $t=1$):
whether Spanish was spoken at home during childhood (``Language$_{t=1}$'') and Hispanic origin (``Origin$_{t=1}$'').
In a subsequent reinterview ($t=2$), two additional indicators of ethnicity were obtained:
 Hispanic ancestry (``Ancestry$_{t=2}$'') and a repetition of the ``Origin''  measure (``Origin$_{t=2}$'').
Following \citet{johnson1990measurement}, we  analyze the group of 9485 respondents not born in an Hispanic country. 
Of interest are false positive rates, $\text{Pr}(Y=\text{Yes}|\xi=\text{No})$, and false negative rates, $\text{Pr}(Y=\text{No}|\xi=\text{Yes})$,  for the alternative question 
formulations. 

\citet{johnson1990measurement} first fitted a two-class model to these data,  yielding a deviance 
 of 103.6 with 6 degrees of freedom ($p < 10^{-5}$), and a Bayesian Information Criterion (BIC) of 48.7.  
The two class model's lack of fit to the data led the authors to subsequently fit a model with two 
separate two-class latent variables corresponding to the two measurement occasions. In terms of probabilities their model can be written
\begin{multline}\label{eq:model-johnson}\text{Pr}(\text{Language$_{t=1}$},\text{Origin$_{t=1}$}, \text{Ancestry$_{t=2}$}, \text{Origin$_{t=2}$}) = \\
\sum_{\{j,k\} \in \{1,2\}\times\{1,2\}} \bigl[	\text{Pr}(\xi_{t=1}=j, \xi_{t=2}=k) 
		\text{Pr}(\text{Language$_{t=1}$} | \xi_{t=1} = j) \text{Pr}(\text{Origin$_{t=1}$} | \xi_{t=1} = j) \bigr. \times\\
		\bigl. \text{Pr}(\text{Ancestry$_{t=2}$} | \xi_{t=2} = k) \text{Pr}(\text{Origin$_{t=2}$} | \xi_{t=2} = k) 
			\bigr].
\end{multline}
It can be seen in Equation \eqref{eq:model-johnson} that instead of one single $\xi$ variable, two latent discrete variables $\xi_{t=1}$ and $\xi_{t=2}$ are defined. Crucially, the conditional probability of an item at a time point only depends on the latent variable corresponding to that time point. Conditional independence given the time-specific latent variable is assumed. The relationship between the two latent variables $\text{Pr}(\xi_{t=1}=j, \xi_{t=2}=k) $ is freely estimated, but could equally well be viewed as a set of four class proportions. An alternative way of viewing this model is therefore as a highly restricted four-class model, where each of the four classes corresponds to a cell in the cross-table of the two latent variables \citep{hagenaars1988latent}. Due to these restrictions the model parameters are identifiable. 

\begin{table}[tb]
\caption{Conditional probability estimates in Johnson's (1990) two-variable model.
Shown are the conditional probabilities for each item given its corresponding latent variable ($\xi_{t=1}$ or $\xi_{t=2}$).}\label{tab:ethnicity-johnson-est}
		\begin{tabular}{rrrrrrrrrrrrr}
	\hline
Class&&\multicolumn{2}{c}{Ancestry$_{t=2}$}&&\multicolumn{2}{c}{Language$_{t=1}$}&&\multicolumn{2}{c}{Origin$_{t=1}$}&&\multicolumn{2}{c}{Origin$_{t=2}$}\\
\cline{3-4}\cline{6-7}\cline{9-10}\cline{12-13}
\#		&& No	& Yes	&& No	& Yes	&& No	& Yes	&& No	& Yes	\\
\hline
1&&0.999	&0.001	&	&0.995	&0.005	&	&0.998	&0.002	&	&0.999	&0.001\\
2&&0.186	&0.814	&	&0.171	&0.829	&	&0.218	&0.782	&	&0.075	&0.925\\
\hline
	\end{tabular}
\end{table}
\citet{johnson1990measurement}'s final analysis is model \eqref{eq:model-johnson} applied to the Hispanic ethnicity data. This indeed improved the deviance to 3.1 with 4 degrees of freedom ($p = 0.54$; BIC equals -33.5). Conditional probability estimates based on the two-variable model are shown in Table \ref{tab:ethnicity-johnson-est}. For conciseness, the probabilities in Table \ref{tab:ethnicity-johnson-est} are conditional on the latent class variable corresponding to the item in question (i.e., $\xi_{t=1}$ or $\xi_{t=2}$).

The latent variables represent
a nuisance dependency due to the measurement occasion (also remarked by the original authors, p. 64). The resulting conditional probabilities from model \eqref{eq:model-johnson} in Table \ref{tab:ethnicity-johnson-est} are therefore difficult to interpret in sociological terms, because we obtain measurement properties of each item as a measurement of the specific time point, $\text{Pr}(Y_{t=1,2} | \xi_{t=1,2})$, but not the false positive and false negative rates of interest, $\text{Pr}(Y_{t} | \xi)$. Instead of the multiple latent variable model \eqref{eq:model-johnson}, it may therefore be preferable to fit a model with a single dichotomous latent class variable but that does account for the time dependence between items.

The question is now whether, starting from the independence model, the score and EPC measures discussed above could have succeeded in detecting the relevant dependencies.
Table \ref{tab:ethnicity} shows the results of calculating the $\EPC_L$, $\EPCgs$, and the corresponding score statistics after fitting the two-class independence model. The dependence between items measured at the same time point is clearly indicated as the primary source of misfit. Moreover, the other pairs of items exhibit negative dependence. Such negative dependence is commonly thought to occur when there is multidimensionality among items measuring different latent variables \citep[e.g.,][p. 127]{yen1984effects}. Here, however, the multidimensionality in question is not of substantive interest. It merely represents extraneous factors of the measurement occasion which are not the focus of the investigation. An alternative to the multidimensional model is therefore a model that frees the two large and positive local dependence parameters, Origin$_{t=2}$	$\leftrightarrow$	Ancestry$_{t=2}$ and Origin$_{t=1}$	$\leftrightarrow$	Language$_{t=1}$. Such a model retains its interpretability as a measurement model for Hispanic ethnicity while also accounting for time dependence.

\begin{table}[tb]\caption{
	Expected parameter changes (EPC's) and score tests (T) for local dependence between indicators of Hispanic ethnicity in the Census.
}\label{tab:ethnicity}
\begin{center}
\begin{tabular}{rrrrrrr}
  \hline
\multicolumn{3}{l}{Local dependence} & $\EPC_L$ & $T_L$ & $\EPCgs$ & $T_{\mathrm{GS}}$ \\
  \hline
  Language$_{t=1}$	&$\leftrightarrow$	&Ancestry$_{t=2}$	&-0.92	&5.0 & -1.45	&	7.9\\
Origin$_{t=1}$	&$\leftrightarrow$	&Ancestry$_{t=2}$	&-1.08	&7.9 & -1.76	&	12.8\\
Origin$_{t=2}$	&$\leftrightarrow$	&Ancestry$_{t=2}$	&4.14	&97.1 & 1.59	&	37.2\\
Origin$_{t=1}$	&$\leftrightarrow$	&Language$_{t=1}$	&2.94	&45.6 & 1.32	&	20.5\\
Origin$_{t=2}$	&$\leftrightarrow$	&Language$_{t=1}$	&-0.76	&2.5 & -1.23	&	4.1\\
Origin$_{t=2}$	&$\leftrightarrow$	&Origin$_{t=1}$	&-1.10	&6.1&-2.20	&	12.2\\
   \hline
\end{tabular}
\end{center}
\end{table}

\begin{table}[tb]
\caption{Conditional probability estimates in final model including local dependencies.}\label{tab:ethnicity-final-est}
		\begin{tabular}{rrrrrrrrrrrrrr}
	\hline
\multicolumn{2}{c}{Class}&&\multicolumn{2}{c}{Ancestry$_{t=2}$}&&\multicolumn{2}{c}{Language$_{t=1}$}&&\multicolumn{2}{c}{Origin$_{t=1}$}&&\multicolumn{2}{c}{Origin$_{t=2}$}\\
\cline{1-2}\cline{4-5}\cline{7-8}\cline{10-11}\cline{13-14}
\#	&Size	&&No	&Yes	&&No	&Yes	&&No	&Yes	&&No	&Yes	\\
\hline
1	&0.976	&&0.999	&0.001	&&0.995	&0.005	&&0.998	&0.002	&&0.999	&0.001\\
2	&0.024	&&0.383	&0.617	&&0.288	&0.712	&&0.329	&0.671	&&0.299	&0.701\\
\hline
	\end{tabular}
\end{table}

Based on the dependencies indicated as substantively relevant in Table \ref{tab:ethnicity}, we fit the loglinear latent class model with two classes and two loglinear local dependencies. The class sizes and conditional probability estimates from this model are shown in Table \ref{tab:ethnicity-final-est}.
Contrary to those in Table \ref{tab:ethnicity-johnson-est}, the conditional probabilities in Table  \ref{tab:ethnicity-final-est} can be interpreted as sensitivity and specificity estimates. This model produces identical expected frequencies and deviance to the multidimensional model
chosen by \citet{johnson1990measurement}, and is therefore equivalent to it. Crucially, however,  the false negative rates of interest differ considerably. Since the nuisance dependencies due to measurement occasions are absorbed by the local dependence parameters $\psi$, the false negative rates can be interpreted as being  with respect to a common latent class variable that might be labeled ``Hispanic ethnicity''.

\begin{table}
\caption{Local dependence estimates in final model including two log-odds local dependencies.}\label{tab:ethnicity-final-dep}
\centering\begin{tabular}{rrrrr}
	\hline
\multicolumn{3}{l}{Local dependence}&	Est.	& Wald\\
	\hline
Origin$_{t=2}$	&$\leftrightarrow$	&Ancestry$_{t=2}$	&	3.166	& 50.2\\
Origin$_{t=1}$	&$\leftrightarrow$	&Language$_{t=1}$	&	1.677 	& 19.3\\
		\hline
\end{tabular}
\end{table}

Table \ref{tab:ethnicity-final-dep} shows the loglinear dependence parameter estimates obtained from the two-class dependence model. The Wald test under this model indicates the same as the score test under the independence model: that these local dependencies differ significantly from zero. These estimates can be interpreted as the log-odds ratio between two items conditional on the latent class. In this particular model, these log-odds ratios are also the marginal log-odds ratios within classes over all possible response patterns. It can be seen that the local dependencies are rather strong, especially at time $t=2$. 

The final model results in Table \ref{tab:ethnicity-final-est} show that Origin$_{t=1}$ and Origin$_{t=2}$ have very similar measurement properties. This appears  reasonable given that we are dealing with the same measure at two different time points. In contrast, Table \ref{tab:ethnicity-johnson-est} shows large differences between the measurement properties of the same question at different time points. The final model, thus, yields results that are easier to interpret than Johnson's (1990) model, but does account for the local dependencies due to measurement occasion. It leads to two new conclusions for the U.S. Census: 1) Considering the false positive and negative rates in Table \ref{tab:ethnicity-final-est}, Origin and Language may be the better measures of ethnicity, where the choice of measurement occasion is inconsequential; 2) the false negative rates in all indicators are considerable, meaning that the number of U.S. residents of Hispanic ethnicity is likely to be underestimated.

\section{Application 2: Dentistry x-ray ratings}\label{sec:dentistry}

\citet{espeland1989using} used latent class modeling to explain 3869 ratings given by five dentists ($y_1$--$y_5$) to x-rays that may show incipient caries (1) or not (0). Each rating is a binary observed variable, and two latent classes represent true caries state. Fitting the two-class local independence model, which we call $M_0$, to these data yields a badly fitting model with a deviance ($L^2$) of 129.9 on 20 degrees of freedom (bootstrap $p < 10^{-10}$) and a BIC of -35.4. These authors then suggested to increase the number of classes to four.

\citet[pp. 804-6]{qu1996random} re-analyzed these data, and argued that the two-class model taking into account local dependencies is easier to interpret than the four class model suggested by \citet{espeland1989using}. \citet[p. 799]{qu1996random} introduced an alternative approach to taking local dependency into account, whereby the dependencies are parameterized as arising from a Gaussian random effect variable, whose effects are allowed to differ over classes. A reformulation of their model can be obtained by modifying Equation \eqref{eq:design-matrices} as
\begin{equation}
\boldsymbol\eta_t = \X_{(Y)} \boldsymbol{\tau} + 
		\X_{(Y\xi_t)} \boldsymbol{\lambda} + \mathbf{b}_t \cdot \eta,
\end{equation}
where $\eta$ is a Gaussian latent factor (random effects) variable, $\eta \sim \text{N}(0, 1)$, the latent variables $\xi$ and $\eta$ are assumed to be independent, and $\mathbf{b}_t$ are vectors of class-specific factor loadings. 
That is, instead of using direct loglinear effect parameters to model the local dependencies, a continuous latent factor on which all items load is posited.
Parameter estimates can be obtained through numerical integration  \citep{vermunt2013technical}.

\citet[p. 805]{qu1996random} fit their random effects model under the restriction that the loadings for $y_1$ and $y_2$, the first two items, are equal, i.e., $\mathbf{b}_{(1,t)} = \mathbf{b}_{(2,t)}$. This model, $M_{\text{Qu}}$, appears to fit the data quite well, with a deviance of 15.8 on 12 degrees of freedom (bootstrap $p = 0.38$). The BIC is equal to -83.4, so that the improvement in model fit appears to somewhat outweigh the added complexity of this model. 

While it appears to fit the data well, the random effects model $M_{\text{Qu}}$ has two problems. First, the maximum-likelihood solution given by \citet{qu1996random} involves boundary estimates that prevent the estimation of asymptotic standard errors. When standard errors are estimated using nonparametric bootstrapping, four (out of eight) factor loadings do not differ statistically significantly from zero. There thus appears to be considerable room for model simplification and additional parsimony. Second, even under the already rather complex $M_{\text{Qu}}$, a score test for the local dependence parameters reveals one strong and statistically significant residual dependence between $y_3$ and $y_4$ ($T_L = 10.5$, $\EPC_L=-1.22$). This indicates that the random effects model cannot completely account for all local dependencies. 

It is in principle possible to formulate a model that includes local dependence parameters as well as a random effect, but it would be preferable to find a more parsimonious model using loglinear local dependence parameters that also fits the data well. We now demonstrate that the $\EPC$ measures together with the score tests can be used to find a model that does not have the disadvantages of the random effects model, but that does account for local dependence while also retaining the easier-to-interpret two-class solution. 

Under the local independence model $M_0$, we calculated $\EPC$ measures and score tests, shown in Table \ref{tab:dentistry}. 
It can be seen that EPC's and score statistics are large for the five bivariate
local dependencies in rows 2, 3, 5, 7, and 9 of the Table. 

Based on the EPC and score test values under $M_0$ in Table \ref{tab:dentistry}, we proceed to formulate a model in which the loglinear local dependence parameters corresponding to rows 2, 3, 5, 7, and 9 of the Table are freed. This model, which we call $M_1$, has 15 degrees of freedom, a deviance of 35.7 ($p = 0.0019$), and a BIC of -88.2. Although the BIC would prefer this model over the random effects model, there still appears to be model misfit. The largest $\EPC_L$ (0.93) and score test ($T_L = 8.8$) are those for the fixed local dependence between $y_1$ and $y_5$ (row 4 in the Table). Moreover, the free local dependence parameter between $y_1$ and $y_4$ (row 3 in the Table) is estimated at a small (0.16) and not statistically significant value ($p = 0.59$) under $M_1$. 

\begin{table}[tb]\caption{EPC and score tests for loglinear local dependence parameters between five dentists' x-ray ratings for caries under the local independence model $M_{0}$.
}\label{tab:dentistry}
\begin{center}  
\begin{tabular}{lcccrrrrr}
  \hline
Number &\multicolumn{4}{l}{Bivariate dependence} & $\EPC_L$ & $T_L$ & $\EPC_{\mathrm{GS}}$  &$T_{\mathrm{GS}}$\\
  \hline
1	& 1	&$\leftrightarrow$&2	&&	0.32	&	3.1	&	0.35	&	3.4	 \\
2	& 1	&$\leftrightarrow$&3	&&	1.04	&	34.0	&	0.97	&	31.6	 \\
3	& 1	&$\leftrightarrow$&4	&&	0.59	&	13.1	&	0.59	&	13.1	 \\
4	& 1	&$\leftrightarrow$&5	&&	0.47	&	2.7	&	0.44	&	2.6	 \\
5	& 2	&$\leftrightarrow$&3	&&	0.56	&	6.8	&	0.53	&	6.4	 \\
6	& 2	&$\leftrightarrow$&4	&&	0.23	&	1.8	&	0.22	&	1.7	 \\
7	& 2	&$\leftrightarrow$&5	&&	0.63	&	16.4	&	0.48	&	12.6	 \\
8	& 3	&$\leftrightarrow$&4	&&	-0.30	&	2.7	&	-0.35	&	3.2	 \\
9	& 3	&$\leftrightarrow$&5	&&	0.76	&	5.1	&	0.55	&	3.7	 \\
10	& 4	&$\leftrightarrow$&5	&&	0.42	&	3.5	&	0.27	&	2.3	 \\
\hline
\end{tabular}
\end{center}
\end{table}


Our final model ($M_2$) therefore fixed the $y_1 \leftrightarrow y_4$ dependence parameter to zero and freed the $y_1 \leftrightarrow y_5$ dependence, as suggested by the EPC measures obtained under $M_1$. The final model $M_2$ has a deviance of 28.4 with 15 degrees of freedom (bootstrap $p = 0.07$) and a  BIC of -95.5. This model therefore appears to fit the data well, and is strongly preferred by BIC over $M_{\text{Qu}}$, $M_{0}$, and $M_{1}$.

Although the local dependence parameters are not of scientific interest in this application, it may aid understanding of the loglinear local dependence model to examine these values. Table \ref{tab:dentists-dependence} provides the estimates of the local dependence parameters under $M_2$, together with their corresponding Wald tests. In addition to these parameter estimates, it is possible to compute the within-class log-odds ratio between a pair of items, marginalized over all other variables (we thank an anonymous reviewer for this suggestion). These marginal log-odds ratios are given in the final two columns of Table \ref{tab:dentists-dependence}. Because the locally dependent item pairs overlap, the $\psi$ parameters no longer correspond to the marginal dependence within classes, as can be seen in the Table. Furthermore, the marginalized local dependencies differ over classes even though the loglinear dependence parameters $\psi$ do not. Finally, it can be seen that the marginalized within-class dependence between $y_1$ and $y_2$ (row 1 of Table \ref{tab:dentists-dependence}) is nonzero even though the loglinear dependence parameter is fixed to zero. This shows that the loglinear dependence model can account for certain, possibly class-dependent, marginal local dependencies in a rather parsimonious manner. 

\begin{table}[tb]
\begin{center}
\caption{Final local dependence latent class model ($M_2$) results. Estimates of log-odds dependencies ($\hat{\psi}$) between pairs of items with Wald tests under $M_2$. Also shown is the implied marginal dependence between pairs of items within classes. }
\label{tab:dentists-dependence}
\begin{tabular}{lccccrrrr}
  \hline
&&&&& \multicolumn{4}{c}{Log-odds ratios within classes}\\
&&&&& \multicolumn{2}{c}{Conditional} & \multicolumn{2}{c}{Marginal}\\
\cline{6-7} \cline{8-9}
Number &\multicolumn{4}{c}{Bivariate dependence} & Est. $(\hat{\psi})$ & Wald & Class 1 & 
	Class 2 \\ 
  \hline
1	& 1&$\leftrightarrow$&2&& - &      &0.25 & 0.68 \\ 
2	& 1&$\leftrightarrow$&3&& 1.38 &49.7  &1.48 & 1.66 \\ 
3	& 1&$\leftrightarrow$&4&& - &      &0 & 0 \\ 
4	& 1&$\leftrightarrow$&5&& 0.74 &60.4  &0.84 & 1.25 \\ 
5	& 2&$\leftrightarrow$&3&& 1.29 &8.0   &1.40 & 1.57 \\ 
6	& 2&$\leftrightarrow$&4&& - &      &0 & 0 \\ 
7	& 2&$\leftrightarrow$&5&& 0.73 &34.4  &0.79 & 1.21 \\ 
8	& 3&$\leftrightarrow$&4&& - &      &0 & 0 \\ 
9	& 3&$\leftrightarrow$&5&& 1.38 &59.0  &1.58 & 1.85 \\ 
10	& 4&$\leftrightarrow$&5&& - &      &0 & 0 \\ 
   \hline
      \multicolumn{7}{l}{-: Fixed to zero.}
\end{tabular}
\end{center}
\end{table}

The estimates of specificity, sensitivity, and prevalence under $M_2$ are given in Table \ref{tab:dentists-sensitivity}. It can be seen that specificity  values are quite high, with few false negatives, except for dentist \#5 (0.68). At the same time,  dentist \#5 appears to have a rather high sensitivity (0.82) of x-ray judgments compared with his or her colleagues, who appear to err on the side of non-detection of caries. The estimated values under $M_2$ in Table \ref{tab:dentists-sensitivity} differ from those under $M_{\text{Qu}}$ especially for dentist \#4: $M_{\text{Qu}}$ would estimate this dentist's sensitivity at 0.68, whereas Table \ref{tab:dentists-sensitivity} shows that under the $M_2$ model this estimate is 0.57. If this model is indeed to be preferred, it would appear that this dentist's performance with regard to false negatives is even worse than was previously thought.

\begin{table}[tb]
\caption{Local dependence latent class model results. Prevalence (class sizes), and sensitivity,  $\text{Pr}(y=\text{Yes}|\xi=2)$, and 
	specificity, $\text{Pr}(y=\text{No}|\xi=1)$, estimates for the five dentists' judgments
of x-rays for caries under $M_2$.}\label{tab:dentists-sensitivity}
\begin{small}
	\begin{tabular}{rrrrrrrrrrrrrrrrr}
	\hline
\multicolumn{2}{c}{Class}&&\multicolumn{2}{c}{Dentist 1}&&\multicolumn{2}{c}{Dentist 2}&&\multicolumn{2}{c}{Dentist 3}&&\multicolumn{2}{c}{Dentist 4}&&\multicolumn{2}{c}{Dentist 5}\\
\cline{1-2}\cline{4-5}\cline{7-8}\cline{10-11}\cline{13-14}\cline{16-17}
\#	&Size	&&No	&Yes	&&No	&Yes	&&No	&Yes	&&No	&Yes	&&No	&Yes\\
\hline
1	&0.79	&&0.99	&0.01	&&0.88	&0.12	&&0.96	&0.04	&&1.00	&0.00	&&0.68	&0.32\\
2	&0.21	&&0.63	&0.37	&&0.39	&0.61	&&0.53	&0.47	&&0.43	&0.57	&&0.18	&0.82\\
\hline
	\end{tabular}\end{small}
\end{table}

\section{Conclusion}

We have shown how the $\EPC_L$ and $\EPCgs$ can aid in the detection of local dependence when the commonly made local independence assumption in latent class analysis of binary data does not hold. The asymptotic and finite sample properties of these measures appear adequate for this purpose. Applications to two real datasets previously analyzed by other authors demonstrated the advantage of this approach in trading off model realism and parsimony, and showed that different and more easily interpretable results can be obtained.

Extensions to polytomous data are possible in our framework by adjusting the relevant design matrices. Unless additional restrictions are imposed, the local dependence parameter for a pair of variables will then become multivariate.  Class-specific and trivariate local dependencies can likewise be accommodated, as can latent class models including covariates. Finally, the $\EPC_L$ and $\EPCgs$ could be  applied to other parameters than local dependencies. For example,
 \citet{glas1999modification} suggested examining item bias
 (direct effects of covariates on response variables) in item response models. 
Based on our findings,  the $\EPC_L$, $\EPCgs$, and 
corresponding score statistics 
have been implemented in the standard latent class modeling software Latent GOLD 5, which allows for the above extensions \citep[pp. 133--4]{vermunt2013technical}. The online supplement provides R code \citep{Rlanguage} for the applications.

Although Section 2 developed the EPC measures for  restrictions in a general maximum-likelihood framework with misspecification, our simulation and applications have been limited to the investigation of class-independent loglinear local dependence parameters with exactly one misspecification. An evaluation of the performance of the EPC measures for such  parameters, as well as in the case of more than one misspecification, remain topics for future investigation.

When conditional dependencies are  among non-overlapping pairs of items, the direct loglinear effect parameters will equal the  marginal local dependencies between pairs of items conditional on the latent class. When pairs of items overlap, this is no longer the case, but marginal dependence could be parameterized using the marginal model formulation of \citet{bartolucci2006class}  and \citet{reboussin2008locally}. This will be advantageous when the goal is to interpret the local dependence parameters, or when the base model from which EPC's are calculated already includes local dependencies itself. Development of score tests and EPC measures for marginal models is therefore another interesting topic for future study.

\appendix

\section{Information matrices, Jacobian, and identification of the locally dependent latent class model}

This appendix defines the information matrices, Jacobian, and outer product matrix for (partially) locally dependent latent class models used in the derivation of the $\EPC$. 
We also provide a theorem giving conditions under which the local dependence parameters
are locally identifiable.

By applying the rules of vector differentiation to model \eqref{eq:loglik},
 the Jacobian of the patternwise likelihood vector with
respect to one of the parameter vectors $\boldsymbol{\tau}, \boldsymbol{\lambda}$, 
or $\boldsymbol{\psi}$ is obtained as
\begin{eqnarray}\label{eq:jacobian}
	\Score_{(.)} = \frac{\partial \log \mathrm{Pr}(\mathbf{Y})}{\partial (.)} =
	\sum^T_{t = 1} [\mathbf{1}' \otimes  \mathrm{Pr}(\xi = t | \mathbf{Y} = \mathbf{y}) \circ (\X_{(.)} - E_R[\X_{(.)}]) ],
\end{eqnarray}
where $\circ$ denotes the elementwise (``Hadamard'') product, the kronecker product 
$\otimes$ here serves to duplicate the posterior probabilities columnwise, $\X_{(.)}$ is the design
matrix corresponding to either $\boldsymbol{\tau}, \boldsymbol{\lambda}$, 
or $\boldsymbol{\psi}$, and $E_R[\X_{(.)}]$ is a matrix with $R$ rows, in which each row equals $\X_{(.)}'\mathrm{Pr}(\mathbf{Y} = \mathbf{y} | \boldsymbol{\xi})$. 
For 
a two-class model with effect coding, the Jacobian with 
respect to the latent class intercept parameter is
\begin{align}
	\Score_{\alpha}= \frac{\partial \log \mathrm{Pr}(\mathbf{Y})}{\partial \alpha} = 2 [\mathrm{Pr}(\xi = 1 | \mathbf{Y} = \mathbf{y}) - \mathrm{Pr}(\xi = 1)].\label{eq:gradient-prior}
\end{align}
That is, the Jacobian depends on the change in the latent class classification
before and after observation of $\mathbf{Y}$. This change therefore plays a large role in the 
determinant of the outer product of the patternwise score vectors used below.

Using obvious notation for the full Jacobian $\Score(\param)$, the gradient ($p$-score vector) over all response patterns will equal 
\begin{equation}
\score = \frac{\partial \ell(\param)}{\partial \param} = \sum^N_{i=1} \frac{\partial \ell_i(\param)}{\partial \param} = 
	 \, \Score(\param)' \mathbf{n}.
\end{equation}
Define the observed and expected information matrices as
\begin{eqnarray}
\I_{Y} &=& -\frac{\partial \score}{\partial \param'} = 
	-\frac{\partial^2 \ell(\param)}{\partial \param \partial\param'},\\
\I_{L} &=&  E_L(\I_{Y}) = 
	\sum^R_{r=1}\hat{n}_r \Score_r(\param)' \Score_r(\param),
\end{eqnarray} where $\hat{n}_r =n \cdot \mathrm{Pr}(\mathbf{Y} = \mathbf{y}_r)$
 and the outer product matrix as
\begin{equation}
\D =  \sum^R_{r=1}n_r \Score_r(\param)' \Score_r(\param).
\end{equation}

The form of the Jacobian in equation \eqref{eq:jacobian} can be used to determine identifiability.
\newtheorem{lemma}{Lemma}
\begin{lemma}\label{lemma}
	Assume that $\Score_{\param_2}$ is of full column rank.
	Let $\mathbf{X}_{\mathrm{new}}$ denote a design matrix such that: 
	\begin{enumerate}
		\item[(i.)] $\mathbf{X}_{\mathrm{new}}$ is of full column rank;
		\item[(ii.)] The number of columns in $\mathbf{X}_{\mathrm{new}}$ is smaller than 
			or equal to the number of degrees of freedom $df = R - 1 - \mathrm{rk}(\Score_{\param_2})$;
		\item[(iii.)] The columns of $\mathbf{X}_{\mathrm{new}}$ are linearly independent of the columns of the design matrix $\mathbf{X}_{\param_2}$ corresponding to the parameters $\param_2$;
		\item[(iv.)] $\mathbf{X}_{\mathrm{new}, t} = \mathbf{X}_{\mathrm{new}}$ for all $t \in \{1..T\}$.
	\end{enumerate}
	Then the parameters $\param_{\mathrm{new}}$ corresponding to $\mathbf{X}_{\mathrm{new}}$ in  model \eqref{eq:loglik} are locally identifiable.
\end{lemma}
\begin{proof}
To show local identifiability, it suffices to show that $\Score_{\mathrm{new}}$ is of full column rank and its columns linearly independent of those in $\Score_{\param_2}$ \citep{goodman1974exploratory}.
	Since $\mathbf{X}_{\mathrm{new}}$ is not class-specific by (iv), equation \eqref{eq:jacobian} reduces to $\Score_{\mathrm{new}} = \mathbf{X}_{\mathrm{new}} - E_R(\mathbf{X}_{\mathrm{new}})$, so  that $\mathrm{rk}(\Score_{\mathrm{new}}) = \mathrm{rk}(\mathbf{X}_{\mathrm{new}})$, implying full column rank by (i). Furthermore, by assumption $\mathrm{rk}(\Score_{\param_2}) = \mathrm{rk}(\mathbf{X}_{\param_2})$, so that by equation \eqref{eq:jacobian}, (ii) and (iii) guarantee that the columns of $\Score_{\mathrm{new}}$ are also independent of those in $\Score_{\param_2}$.
\end{proof}
The proof of Theorem 1 follows directly from the Lemma and model \eqref{eq:design-matrices}. It suggests that when the local independence model is identifiable and the number of local dependencies
$\psi$ to be freed does not exceed the degrees of freedom, these additional parameters will also be identifiable.

\section{Supplemental materials}

\begin{description}
\item[R code:] Provides S4 classes to perform latent class analysis for binary variables with local dependencies and obtain the $\EPC_L$ and $\EPCgs$ and score tests. Includes both data sets used as examples in the article. (GNU zipped tar file)
\end{description}

\bibliography{/Users/daob/Dropbox/Bibliography/quality}

\end{document}